\documentstyle[amssymb,preprint,aps]{revtex}

\begin{document}

\begin{center}
{\huge Teleportation of a Zero- and One-photon Running Wave State by
Projection Synthesis} \\[0.25cm]
{\bf C. J. Villas-B\^{o}as, N. G. Almeida, and } {\bf M. H. Y. Moussa}$^{*}$%
{\bf \ }\\[0pt]
{\it Departamento de F\'{\i }sica, Universidade Federal de S\~{a}o Carlos,\\[%
0pt]
Via Washington Luis, km 235, S\~{a}o Carlos 13565-905, SP, Brasil}\\[0pt]
\end{center}


\begin{center}
{\bf Abstract}
\end{center}

We show how to teleport a running wave superposition of zero- and one-photon
field state through the projection synthesis technique. The fidelity of the
scheme is computed taking into account the noise introduced by dissipation
and the efficiency of the detectors. These error sources have been
introduced through a single general relationship between input and output
operators.

\noindent PACS number(s): \ 03.65.Bz, 03.67.-a, 42.50.Dv

\noindent $^*${\it E-mail: miled@power.ufscar.br}\newline

About five years after the proposition of quantum teleportation by Bennett 
{\it et. al.} \cite{bennett}, this phenomenon has recently been demonstrated
in a couple of experiment \cite{dik,boschi} through photon polarized states.
The required quantum channel, an EPR state \cite{epr}, has been generated,
in both experimental realizations, by photons emerging from a type II
degenerate parametric down-conversion \cite{kwiat}. Basically, a
superposition of horizontally ($\left| h\right\rangle $) and vertically ($%
\left| v\right\rangle $) polarized states of photon $1$, {\it i.e.}, $\alpha
\left| h\right\rangle +\beta \left| v\right\rangle $, is teleported to
another photon, say $3$, which is part of an entangled quantum channel $%
(\left| v\right\rangle \left| h\right\rangle +\left| h\right\rangle \left|
v\right\rangle )/\sqrt{2}$.

Despite the fact that the experimental demonstration of teleportation of an
atomic state has not yet been realized, there is a number of proposals on
this subject \cite{davi,norton}. There is even a proposal that considers two
particles whose states are simultaneously teleported one to the other, the
identity interchange process \cite{miled2}. Experimental proposals for
teleporting a Schr\"{o}dinger cat state of the radiation field, both trapped
in a high-${\em Q}$ cavity \cite{miled3} and as a running wave \cite{kimble}%
, have been reported. In Ref. \cite{kimble} the authors analyse
teleportation of continuous quantum variables and calculate the fidelity of
the process. Teleportation of $N$-dimensional states has also been proposed
in the cavity QED domain and regarding other physical systems \cite{miled1}.

In the present paper we show, for the first time, how to teleport a running
wave superposition of zero- and one-photon field states. The teleportation
machine is based on a recently reported scheme for optical state truncation
by projection synthesis \cite{barnett}. Such a ``quantum-scissors'' device
is suitable for teleportation since both the processes rely exactly on the
same principles: the possibility of entanglement, and so, nonlocality, and
the projection postulate. The technique of projection synthesis has been
proposed originally by Barnett and Pegg \cite{barnett1} for the experimental
determination of the canonical quantum optical phase probability
distribution, and has also been applied for ${\em Q}$-function measurement 
\cite{basilio}.

As sketched in Fig. 1a, the teleportation experiment consists of a couple of
50/50 symmetric beam splitters, $BS_1$ and $BS_2$, and a couple of
photodetectors, $D_b$ and $D_c$. As in the original Bennett's teleportation
scheme, Bob is supposed to share a quantum channel with Alice, here an
entanglement composed by the output fields $a$ and $b$. While the output
field $a$, which is expected to receive the teleported state, is given to
Bob, the output field $b$ is given to Alice. She is expected to realize a
Bell-type measurement on field $b$ itself and the one injected through the
input mode $c$ in the state to be teleported. Obviously, such a state in
mode $c$ is supposedly unknown to both Alice and Bob. As the output modes $a$
and $b$ had been previously entangled, the phenomenon of nonlocality plus
the projection postulate lead to the achievement of the teleportation
process \cite{bennett}.

{\it Ideal process}. Let us briefly consider the situation where no losses
are introduced. The state to be teleported, injected through the input mode $%
c$ and expected to be described as

\begin{equation}
\left| \psi \right\rangle _c=c_0\left| 0\right\rangle _c+c_1\left|
1\right\rangle _c,  \label{1}
\end{equation}
is prepared through a quantum scissors device \cite{barnett}, which is a
replica of the teleportation machine as depicted in Fig.1b. To prepare the
state (\ref{1}) a single-photon field is injected through the input mode $c$
on $BS_1$ resulting in the entangled field $\left| \psi \right\rangle _{cd}=%
\frac 1{\sqrt{2}}\left( \left| 10\right\rangle _{cd}+i\left| 01\right\rangle
_{cd}\right) $. Next, a coherent field $\left| \gamma \right\rangle
_e=\sum_{n=0}^\infty $ $\gamma _n\left| n\right\rangle _e$ $=\gamma _0\left|
0\right\rangle _e+\gamma _1\left| 1\right\rangle _e+...$, is sent through
the input mode $e$ with its vacuum and one photon amplitudes satisfying $%
\gamma _0/c_0=\gamma _1/c_1={\cal C}$ $\left( \left| {\cal C}\right|
^2=\left( \left| \gamma _0\right| ^2+\left| \gamma _1\right| ^2\right)
\right) $. As shown in \cite{barnett}, when registering a single photon in
detector $D_d$ and no counts in $D_e$, one synthesizes the projection of the
entanglement resulting from $BS_1^{\prime }$ on a given state of the output
field $d$, {\it i.e.}, $_d\langle \phi |\psi \rangle _{cd}$, leading to the
prepared state (\ref{1}). Thus, when also accounting for the preparation of
the state to be teleported, the whole machine consists of a double
quantum-scissors device, the output mode $c$ of that suppose to prepare the
state to be teleported (Fig.1b) being the input mode $c$ of the
teleportation device (Fig.1a).

In Ref.\cite{barnett} the authors have pointed out that the quantum-scissors
device is a limited form of teleportation in that part of the coherent state 
$\left| \gamma \right\rangle _e$, formed from the vacuum and one-photon
states, is ``teleported'' to mode $c$. However, we show below that it is
possible to teleport a given quantum superposition of zero- and one-photon
field state from mode $c$ to mode $a$.

Simultaneously to the preparation of the state to be teleported, the quantum
channel is prepared through a single-photon field $a$ incident on $BS_1$ in
a way to superpose on $BS_2$ both the state to be teleported and the output
field $b$ entangled to $a$ as $\left| \psi \right\rangle _{ab}=\frac 1{\sqrt{%
2}}\left( \left| 10\right\rangle _{ab}+i\left| 01\right\rangle _{ab}\right) $
. The product of Alice's state to be teleported and the quantum channel can
be expanded, apart from an irrelevant phase factor, as 
\begin{eqnarray}
&&\frac 12\left[ \left| \Psi ^{-}\right\rangle _{bc}\left( c_0\left|
0\right\rangle _a+c_1\left| 1\right\rangle _a\right) +\left| \Phi
^{-}\right\rangle _{bc}\left( c_1\left| 0\right\rangle _a+c_0\left|
1\right\rangle _a\right) -\right.  \nonumber \\
&&\left. \left| \Psi ^{+}\right\rangle _{bc}\left( c_0\left| 0\right\rangle
_a-c_1\left| 1\right\rangle _a\right) -\left| \Phi ^{+}\right\rangle
_{bc}\left( c_1\left| 0\right\rangle _c-c_0\left| 1\right\rangle _c\right) 
\right] ,  \label{tel}
\end{eqnarray}
where we have introduced the complete set of eigenstates of Bell operators

\begin{mathletters}
\begin{eqnarray}
\left| \Psi ^{\pm }\right\rangle _{bc} &=&\frac 1{\sqrt{2}}\left( \left|
01\right\rangle _{bc}\pm i\left| 10\right\rangle _{bc}\right) ,
\label{bella} \\
\left| \Phi ^{\pm }\right\rangle _{bc} &=&\frac 1{\sqrt{2}}\left( \left|
00\right\rangle _{bc}\pm i\left| 11\right\rangle _{bc}\right) .
\label{bellb}
\end{eqnarray}

Hence, a measurement using Bell states analysers on fields $b$ and $c$
projects the mode $a$ on a superposition of zero- and one-photon field state
as described in (\ref{tel}). This required joint measurement can be achieved
through $BS_2$ by the projection synthesis technique.. In fact, by
superposing the field state to be teleported with the output field $b$ in $%
BS_2$, we get

\end{mathletters}
\begin{mathletters}
\begin{eqnarray}
\widehat{U}\left| \Psi ^{\pm }\right\rangle _{bc} &\varpropto &\left\{ 
\begin{array}{r@{\quad\quad}l}
\left| 01\right\rangle _{bc} &  \\ 
\left| 10\right\rangle _{bc} & 
\end{array}
\right. ,  \label{meda} \\
\widehat{U}\left| \Phi ^{\pm }\right\rangle _{bc} &\varpropto &\left\{
\left| 00\right\rangle _{bc}\mp \frac 1{\sqrt{2}}\left( \left|
20\right\rangle _{bc}+\left| 02\right\rangle _{bc}\right) \right. ,
\label{medb}
\end{eqnarray}

where $\widehat{U}=\exp \left[ i\frac \pi 4\left( \widehat{b}^{\dagger }%
\widehat{c}+\widehat{c}^{\dagger }\widehat{b}\right) \right] $ is the
unitary operator describing the action of an ideal $BS_2$. We thus see from
Eqs. (\ref{meda}) and (\ref{medb}) that a measurement of the field state $%
\left| 10\right\rangle _{bc}$, which requires the incoming Bell state $%
\left| \Psi ^{-}\right\rangle _{bc}$, projects the output field $a$ exactly
on the original state of field $c$. Otherwise, a joint measurement of the
Bell state $\left| \Psi ^{+}\right\rangle _{bc}$ is achieved by measuring
the field state $\left| 10\right\rangle _{bc}$ leaving the output field $a$
in the original state $\left| \psi \right\rangle _c$ but phase shifted
through $\pi $. However, the whole process, beginning with the preparation
of the state to be teleported will be developed below in a noise environment
and taking account of detector inefficiency.

{\it Losses in the }$BS${\it 's. }With the inclusion of errors due to
photoabsortion in the beam splitters, the general relationships between the
input and output operators $\widehat{\alpha },\widehat{\beta }$,{\it \ i.e.}%
, $\widehat{a},\widehat{b}$ ($\widehat{d},\widehat{c}$) in $BS_1$ ($%
BS_1^{\prime }$) or $\widehat{b},\widehat{c}$ ($\widehat{d},\widehat{e}$) in 
$BS_{2\text{ }}$($BS_{2\text{ }}^{\prime }$), are \cite{jeffers}

\end{mathletters}
\begin{mathletters}
\begin{eqnarray}
\widehat{\alpha }_{out} &=&t\widehat{\alpha }_{in}+r\widehat{\beta }_{in}+%
\widehat{{\cal L}}_\alpha ,  \label{2a} \\
\widehat{\beta }_{out} &=&t\widehat{\beta }_{in}+r\widehat{\alpha }_{in}+%
\widehat{{\cal L}}_\beta ,  \label{2b}
\end{eqnarray}

where $t$ and $r$ are the beam-splitter transmission and reflection
coefficients, respectively. In fact, such coefficients, and so the
operators, depend on the frequency of the fields and here a monochromatic
source is considered. The input fields and the noise sources are required to
be independent so that the input operators must commute with the output
Langevin operators:

\end{mathletters}
\begin{equation}
\left[ \widehat{\alpha }_{in},\widehat{{\cal L}}_\alpha \right] =\left[ 
\widehat{\alpha }_{in},\widehat{{\cal L}}_\beta \right] =\left[ \widehat{%
\alpha }_{in},\widehat{{\cal L}}_\alpha ^{\dagger }\right] =\left[ \widehat{%
\alpha }_{in},\widehat{{\cal L}}_\beta ^{\dagger }\right] =0,  \label{3}
\end{equation}
with similar relations for the $\beta $ operators. Imposition of the bosonic
commutation relations on the output mode operators then leads to the
requirements on the noise-operator commutation relations:

\begin{mathletters}
\begin{eqnarray}
\left[ \widehat{{\cal L}}_\alpha ,\widehat{{\cal L}}_\alpha ^{\dagger }%
\right] &=&\left[ \widehat{{\cal L}}_\beta ,\widehat{{\cal L}}_\beta
^{\dagger }\right] =\Gamma ,  \label{4a} \\
\left[ \widehat{{\cal L}}_\alpha ,\widehat{{\cal L}}_\beta ^{\dagger }\right]
&=&\left[ \widehat{{\cal L}}_\beta ,\widehat{{\cal L}}_\alpha ^{\dagger }%
\right] =-\Omega ,  \label{4b}
\end{eqnarray}

where $\Gamma =1-\left| t\right| ^2-\left| r\right| ^2$ is the damping
constant and $\Omega =tr^{*}+rt^{*}$. For optical frequencies the state of
the environment can be very well approximated by the vacuum state even at
room temperature, so that 
\end{mathletters}
\begin{equation}
\widehat{{\cal L}}_\alpha \left| 0\right\rangle =\widehat{{\cal L}}_\beta
\left| 0\right\rangle =\widehat{\alpha }_{in}\left| 0\right\rangle =\widehat{%
\beta }_{in}\left| 0\right\rangle =0,  \label{5}
\end{equation}
and, from the input-output relations (\ref{2a} and \ref{2b}), it also
follows that

\begin{equation}
\widehat{\alpha }_{out}\left| 0\right\rangle =\widehat{\beta }_{out}\left|
0\right\rangle =0.  \label{6}
\end{equation}
Finally, the quantum averages of the Langevin operators vanish, 
\begin{equation}
\left\langle \widehat{{\cal L}}_\alpha \right\rangle =\left\langle \widehat{%
{\cal L}}_\beta \right\rangle =\left\langle \widehat{{\cal L}}_\alpha
^{\dagger }\right\rangle =\left\langle \widehat{{\cal L}}_\beta ^{\dagger
}\right\rangle =0,  \label{7}
\end{equation}
and the only nonzero ground-state expectation values for the products of
pairs of noise operators are

\begin{mathletters}
\begin{eqnarray}
\left\langle \widehat{{\cal L}}_\alpha \widehat{{\cal L}}_\alpha ^{\dagger
}\right\rangle &=&\left\langle \widehat{{\cal L}}_\beta \widehat{{\cal L}}%
_\beta ^{\dagger }\right\rangle =\Gamma ,  \label{8a} \\
\left\langle \widehat{{\cal L}}_\alpha \widehat{{\cal L}}_\beta ^{\dagger
}\right\rangle &=&\left\langle \widehat{{\cal L}}_\beta \widehat{{\cal L}}%
_\alpha ^{\dagger }\right\rangle =-\Omega .  \label{8b}
\end{eqnarray}

As noted in Ref. \cite{jeffers}, the above relations for the averages of the
Langevin operators may also be derived from a canonical one-dimensional
theory applied to a dielectric slab.

Next, it is easy to conclude that, similar to the relations (\ref{2a}) and (%
\ref{2b}), the transformation leading from the output to the input operators
preserving the above-mentioned properties for the Langevin operators read

\end{mathletters}
\begin{mathletters}
\begin{eqnarray}
\widehat{\alpha }_{in} &=&t^{*}\widehat{\alpha }_{out}+r^{*}\widehat{\beta }%
_{out}+\widehat{{\cal L}}_\alpha ,  \label{9a} \\
\widehat{\beta }_{in} &=&t^{*}\widehat{\beta }_{out}+r^{*}\widehat{\alpha }%
_{out}+\widehat{{\cal L}}_\beta ,  \label{9b}
\end{eqnarray}

where the bosonic commutation relations on the input mode operators are
satisfied.

{\it Efficiency of the detectors. }To deal with the efficiency of the
detectors we again take advantage of the Langevin operators. Introducing
output operators accounting for the detection of a given input field $\alpha 
$ (modes $b$,$c$ ($d$,$e$) reaching the detectors in Fig.1a (Fig.1b)), we
write

\end{mathletters}
\begin{equation}
\widehat{\alpha }_{out}=\sqrt{\eta }\widehat{\alpha }_{in}+\widehat{{\frak L}%
}_\alpha ,  \label{90}
\end{equation}
considering the case in which the detectors have the same efficiency $\eta $%
. Obviously, different from the $BS^{\prime }$s the detectors do not couple
different modes in a way that the Langevin operators $\widehat{{\frak L}}%
_\alpha $, despite satisfying all the properties of those introduced above,
obey the commutation relations

\begin{mathletters}
\begin{eqnarray}
\left[ \widehat{{\frak L}}_\alpha ,\widehat{{\frak L}}_\alpha ^{\dagger }%
\right] &=&1-\eta ,  \label{91a} \\
\left[ \widehat{{\frak L}}_\alpha ,\widehat{{\frak L}}_\beta ^{\dagger }%
\right] &=&0,  \label{91b}
\end{eqnarray}
and the ground-state expectation values for the products of pairs are

\end{mathletters}
\begin{mathletters}
\begin{eqnarray}
\left\langle \widehat{{\frak L}}_\alpha \widehat{{\frak L}}_\alpha ^{\dagger
}\right\rangle &=&1-\eta ,  \label{92a} \\
\left\langle \widehat{{\frak L}}_\alpha \widehat{{\frak L}}_\beta ^{\dagger
}\right\rangle &=&0.  \label{92b}
\end{eqnarray}

{\it General relations for the errors. }We next introduce an algebra
accounting for both the errors sources due to photoabsortion in the $%
BS^{\prime }$s (Eqs. (\ref{2a}) and (\ref{2b})) and the efficiency of
detectors (Eq. (\ref{90})). One can check that in such an algebra the output
operators $\widehat{\alpha },\widehat{\beta }$, those describing the fields
reaching the detectors $\widehat{b},\widehat{c}$ ($\widehat{d},\widehat{e}$)
in $BS_{2\text{ }}$($BS_{2\text{ }}^{\prime }$), are

\end{mathletters}
\begin{mathletters}
\begin{eqnarray}
\widehat{\alpha }_{out} &=&{\bf t}\widehat{\alpha }_{in}+{\bf r}\widehat{%
\beta }_{in}+\widehat{{\bf L}}_\alpha ,  \label{93a} \\
\widehat{\beta }_{out} &=&{\bf t}\widehat{\beta }_{in}+{\bf r}\widehat{%
\alpha }_{in}+\widehat{{\bf L}}_\beta ,  \label{93b}
\end{eqnarray}
where ${\bf t=}\sqrt{\eta }t$, ${\bf r=}\sqrt{\eta }r$, and $\widehat{{\bf L}%
}_\alpha =\widehat{{\cal L}}_\alpha +\widehat{{\frak L}}_\alpha $. In fact,
from all the above-mentioned properties of the operators in relations (\ref
{93a}) and (\ref{93b}), we get

\end{mathletters}
\begin{mathletters}
\begin{eqnarray}
\left[ \widehat{{\bf L}}_\alpha ,\widehat{{\bf L}}_\alpha ^{\dagger }\right]
&=&\left[ \widehat{{\bf L}}_\beta ,\widehat{{\bf L}}_\beta ^{\dagger }\right]
=\eta \Gamma +\left( 1-\eta \right) ,  \label{94a} \\
\left[ \widehat{{\bf L}}_\alpha ,\widehat{{\bf L}}_\beta ^{\dagger }\right]
&=&\left[ \widehat{{\bf L}}_\beta ,\widehat{{\bf L}}_\alpha ^{\dagger }%
\right] =-\eta \Omega .  \label{94b}
\end{eqnarray}

When considering $\eta =1$ in (\ref{94a}) and (\ref{94b}) we recover the
relations (\ref{4a}) and (\ref{4b}), while for $\Gamma =0$, which also
implies $\Omega =0$, we recover the relations (\ref{92a}) and (\ref{92b}),
respectively.

{\it Engineering the state to be teleported}. Back to the apparatus in Fig.
1b, when engineering the state to be teleported in a noise environment{\it \ 
}by sending a single-photon field $c$ on $BS_1^{\prime }$, Eq. (\ref{9a})
leads the $c$ and $d$ output fields, together with the environment, in the
entanglement

\end{mathletters}
\begin{equation}
\left( t\left| 10\right\rangle _{cd}+r\left| 01\right\rangle _{cd}+\left|
00\right\rangle _{cd}\widehat{{\cal L}}_c\right) \left| {\bf 0}\right\rangle
_{{\bf E}}.  \label{10}
\end{equation}
Next, on $BS_2^{\prime }$ the field in mode $d$ is coupled to an additional
field in mode $e$ prepared in a coherent state $\left| \gamma \right\rangle
_e=\sum_n\gamma _n\left| n\right\rangle _e$. As above-mentioned the
synthesized projection onto the state in mode $d$ leading to the engineered
state (\ref{1}) results when a single photon is registered in $D_d$ and no
counts in $D_e$, in a way that the output state $c$ plus environment reads

\begin{equation}
_d\left\langle 1\right| _c\left\langle 0\right| \left( t\left|
1\right\rangle _c+r\left| 0\right\rangle _c\widehat{d}_{in}^{\dagger
}+\left| 0\right\rangle _c\widehat{{\cal L}}_c^{\dagger }\right)
\sum_n\gamma _n\frac{\widehat{e}_{in}^{\dagger ^n}}{\sqrt{n!}}\left|
00\right\rangle _{de}{\bf |0\rangle }_{{\bf E}}.  \label{11}
\end{equation}
We note that the environmental states due to both beam splitters have been
put together. Since all the output operators composing $\widehat{e}%
_{in}^{\dagger }$ commute to each other, using the binomial formula

\begin{equation}
\widehat{e}_{in}^{\dagger ^n}=\sum_{k=o}^n\sum_{l=0}^{n-k}%
{n \choose k}%
{n-k \choose l}%
{\bf t}^k{\bf r}^l\widehat{d}_{out}^{\dagger ^l}\widehat{e}_{out}^{\dagger
^k}\widehat{{\bf L}}_e^{\dagger ^{n-k-l}},  \label{110}
\end{equation}
we end up with

\begin{equation}
\left| \psi \right\rangle _{c{\bf E}}={\cal N}\left[ \gamma _0\left|
0\right\rangle _c\left| {\bf \Lambda }_0(t,r)\right\rangle _{{\bf E}}+\gamma
_1\left| 1\right\rangle _c\left| {\bf \Lambda }_1(t,r)\right\rangle _{{\bf E}%
}\right] ,  \label{12}
\end{equation}
where the environmental states read

\begin{mathletters}
\begin{eqnarray}
\left| {\bf \Lambda }_0(t,r)\right\rangle _{{\bf E}} &=&\frac{{\bf r}}{%
\gamma _0}\sum_n\frac{\gamma _n}{\sqrt{n!}}\left( t\widehat{{\bf L}}%
_e^{\dagger }+nr\widehat{{\bf L}}_d^{\dagger }+n\widehat{{\cal L}}%
_c^{\dagger }\right) \widehat{{\bf L}}_e^{\dagger ^{n-1}}{\bf |0\rangle }_{%
{\bf E}},  \label{13a} \\
\left| {\bf \Lambda }_1(t,r)\right\rangle _{{\bf E}} &=&\frac{t{\bf r}}{%
\gamma _1}\sum_n\frac{n\gamma _n}{\sqrt{n!}}\widehat{{\bf L}}_e^{\dagger
^{n-1}}{\bf |0\rangle }_{{\bf E}},  \label{13b}
\end{eqnarray}

Now we use the Wick's theorem for boson operators,

\end{mathletters}
\begin{equation}
\left\langle \widehat{{\bf L}}_\alpha ^{^n}\widehat{{\bf L}}_\beta ^{\dagger
^m}\right\rangle =\delta _{\alpha \beta }\delta _{nm}n!\left\langle \widehat{%
{\bf L}}_\alpha \widehat{{\bf L}}_\beta ^{\dagger }\right\rangle ^n=\delta
_{\alpha \beta }\delta _{nm}n!\left[ \eta \Gamma +\left( 1-\eta \right) %
\right] ^n,  \label{14}
\end{equation}
where for $\eta =1$ it follows that $\widehat{{\bf L}}_\alpha =\widehat{%
{\cal L}}_\alpha $ and for $\Gamma =0$, $\widehat{{\bf L}}_\alpha =\widehat{%
{\frak L}}_\alpha $. From relation (\ref{14}) we get, for the normalization
factor ${\cal N}$ in (\ref{12}), the result

\begin{equation}
{\cal N}=\left\{ 
\mathop{\rm e}%
\nolimits^{\left( \eta \Gamma +(1-\eta )\right) \left| \alpha \right|
^2}\eta \left| r\right| ^2\left[ \left| {\cal C}\right| ^2\left| t\right|
^2+\left( \eta \Gamma +\frac \Gamma {\left| r\right| ^2}+\left( 1-\eta
\right) \right) \left| r\right| ^2\left| \gamma _1\right| ^2\right] \right\}
^{-1}.  \label{15}
\end{equation}
The fidelity of the optical state truncation scheme leading to the
engineered field $\left| \psi \right\rangle _{c{\bf E}}$ , expected to be $%
\left| \Psi \right\rangle _c=\frac 1{{\cal C}}\left( \gamma _0\left|
0\right\rangle _c+\gamma _1\left| 1\right\rangle _c\right) $ when 50/50 $BS^{%
\text{'}}$s are considered ({\it i.e.}, $\left| t\right| =\left| r\right|
=\left| \xi \right| )$, also results from Wick's theorem as

\begin{equation}
{\cal F}=\left\| _c\langle \Psi |\psi \rangle _{c{\bf E}}\right\| ^2=1-\frac{%
1-\eta \left( \frac{1+\Gamma ^2}{1-\Gamma }\right) }{\left( 1+{\cal R}%
\right) \left\{ 1+{\cal R}\left[ 1-\eta \left( \frac{1+\Gamma ^2}{1-\Gamma }%
\right) \right] \right\} },  \label{16}
\end{equation}
where ${\cal R=}\left( \left| \gamma _0\right| /\left| \gamma _1\right|
\right) ^2$. As expected, when considering ideal detectors ($\eta =1$) and
disregarding the losses in the $BS$'s ($\Gamma =0$), we find ${\cal F}=1$.
Moreover, for finite $\eta $ and $\Gamma $, the largest the ratio ${\cal R}$%
, making the probability to find a photon negligible, the closest to unity
is the fidelity. We note that, unlike other situations, when measuring $%
\left| 10\right\rangle _{bc},$ as we have done above, we do not need to
consider 50/50 $BS$'s in order for the relation $\gamma _0/c_0=\gamma _1/c_1=%
{\cal C}$ to be required.

{\it Teleportation process. }As mentioned above,{\it \ }simultaneously to
the preparation of the state to be teleported, which is given to Alice, the
quantum channel has to be prepared by sending a one-photon field state
through $BS_1$. The quantum channel is exactly described by Eq. (\ref{10}),
except that we must change the output symbols $c$ and $d$ by $a$ and $b$,
respectively, while the state to be teleported comprehends the Eq. (\ref{12}%
), as indicated in Fig. 1. For simplicity the state to be teleported will be
rewritten as

\begin{equation}
\left| \psi \right\rangle _{c{\bf E}}={\cal N}\left( \left| \gamma
_0\right\rangle _{{\bf E}}\left| 0\right\rangle _c+\left| \gamma
_1\right\rangle _{{\bf E}}\left| 1\right\rangle _c\right) ,  \label{17}
\end{equation}
with $\left| \gamma _0\right\rangle _{{\bf E}}=\gamma _0\left| {\bf \Lambda }%
_0(t,r)\right\rangle _{{\bf E}}$ and $\left| \gamma _1\right\rangle _{{\bf E}%
}=\gamma _1\left| {\bf \Lambda }_1(t,r)\right\rangle _{{\bf E}}$ . Alice is
thus supposed to realize the joint measurement on fields $b$ and $c$, which
is accomplished through $BS_2$ following exactly the steps outlined in Eqs. (%
\ref{11}-\ref{13b}), substituting the coherent state $|\gamma \rangle _c$ by 
$\left| \psi \right\rangle _{c{\bf E}}$ and the modes $d$, $e$ by $b$, $c$,
respectively. As we see from Eq. (\ref{meda}), in the ideal situation, when
projecting the correlated output fields $b$ and $c$ on the state $\left|
10\right\rangle _{bc}$, as done in Eq. (\ref{11}), we are proceeding to a
Bell measurement of the incoming state $\left| \Psi ^{-}\right\rangle _{bc}$%
. Obviously, this is not the case when the loss mechanisms in the beam
splitters are taken into account. After a straightforward calculation we
obtain, for the teleported field $a,$ the state 
\begin{equation}
\left| \psi \right\rangle _{a{\bf E}}={\sf N}\left( \left| \lambda
_0\right\rangle _{{\bf E}}\left| 0\right\rangle _c+\left| \lambda
_1\right\rangle _{{\bf E}}\left| 1\right\rangle _c\right) ,  \label{18}
\end{equation}
with the normalization constant

\begin{equation}
{\sf N}=\left\{ 
\mathop{\rm e}%
\nolimits^{-\eta (1-\Gamma )\left| \alpha \right| ^2}\eta \left( \frac{%
1-\Gamma }2\right) ^2\left[ 1+\frac 1{{\cal R}}\left( \frac 4{1-\Gamma }%
-3\eta (1-\Gamma )\right) \right] \right\} ^{-1}  \label{180}
\end{equation}
and

\begin{mathletters}
\begin{eqnarray}
\left| \lambda _0\right\rangle _{{\bf E}} &=&\left[ \left| \gamma
_0\right\rangle _{{\bf E}}+\left| \gamma _1\right\rangle _{{\bf E}}\left( 
\widehat{{\bf L}}_c^{\dagger }+\frac rt\widehat{{\bf L}}_b^{\dagger }+\frac 1%
t\widehat{{\bf L}}_a^{\dagger }\right) \right] {\bf |0\rangle }_{{\bf E}},
\label{19a} \\
\left| \lambda _1\right\rangle _{{\bf E}} &=&\left| \gamma _1\right\rangle _{%
{\bf E}}{\bf |0\rangle }_{{\bf E}}.  \label{19b}
\end{eqnarray}

The fidelity of the teleported state expected to be $\left| \Psi
\right\rangle _a=\frac 1{{\cal C}}\left( \gamma _0\left| 0\right\rangle
_a+\gamma _1\left| 1\right\rangle _a\right) $, after computing losses from
both the state engineering scheme and the teleportation process, results

\end{mathletters}
\begin{equation}
\digamma {\em =}\left\| _a\langle \Psi |\psi \rangle _{a{\bf E}}\right\|
^2=1-\frac{\frac{3+\Gamma }{1-\Gamma }-3\eta (1-\Gamma )}{\left( 1+{\cal R}%
\right) \left\{ 1+{\cal R}\left[ \frac 4{1-\Gamma }-3\eta (1-\Gamma )\right]
\right\} },  \label{20}
\end{equation}
where 50/50 $BS$'s were considered. Evidently, for the ideal case $\digamma
=1$ and also the largest ${\cal R}$ the closest to unity is the fidelity.

It is straightforward to obtain the density matrix for both prepared (\ref
{12}) and teleported field state (\ref{17}) by getting rid of the
environmental degrees of freedom. By comparing the prepared field state
density matrix when $\Gamma =0$ we obtain exactly the result presented in
Ref. \cite{barnett}. For estimating the fidelities in Eqs. (\ref{16}) and (%
\ref{20}) we note that the efficiency for single-photon detectors is about $%
70\%$, while the damping constant for $BS^{\text{'}}$s is considerably
small, less than $2\%$ in $BK7$ crystals. As far as we know, until the
present date there is no other scheme for teleportation of a zero- and
one-photon running wave superposition. The present scheme becomes possible
due to the recent proposals for state truncation of traveling optical fields 
\cite{barnett,dakna}. Finally, we note that it is worth proceeding to the
generalization of the present scheme, for teleportation of $N$-dimensional
states, possibly through the Dakna {\it et.al. }engineering technique for
arbitrary quantum state of traveling fields \cite{dakna}.

\begin{center}
{\bf Acknowledgments}
\end{center}

We wish to thank the support from CNPq and FAPESP, Brazil, and R. J.
Napolitano, J. C. Xavier, and P. Nussenzveig for helpful discussions.

\noindent{\bf Figure Caption}

FIG. 1. Sketch of the experimental setup for teleportation by projection
synthesis.


\begin{references}
\bibitem{bennett}  C. H. Bennett, G. Brassard, C. Cr\'{e}peau, R. Jozsa, A.
Peres, and W. Wootters, Phys. Rev. Lett. {\bf 70}, 1895 (1993).

\bibitem{dik}  D. Bouwmeester, J.-W. Pan, K. Mattle, M. Eibl, H. Weinfurter,
and A. Zeilinger, Nature {\bf 390}, 575 (1997).

\bibitem{boschi}  D. Boschi, S. Branca, F. De Martini, L. Hardy, and S.
Popescu, Phys. Rev. Lett. {\bf 80}, 1121 (1998).

\bibitem{epr}  A. Einstein, B. Podolsky, and N. Rosen, Phys. Rev. {\bf 47},
777 (1935).

\bibitem{kwiat}  P. G. Kwiat, K. Mattle, H. Weinfurter, and A. Zeilinger,
Phys. Rev. Lett. {\bf 75}, 4337 (1995).

\bibitem{davi}  L. Davidovich, N. Zagury, M. Brune, J. M. Raimond, and S.
Haroche, Phys. Rev. A {\bf 50}, R895 (1994).

\bibitem{norton}  N. G. Almeida, L. P. Maia, C. J. Villas-B\^{o}as, and M.
H. Y. Moussa, Phys. Lett. A {\bf 241}, 213 (1998).

\bibitem{miled2}  M. H. Y. Moussa, Phys. Rev. A {\bf 55}, R3287 (1997); L.
Vaidman and N. Yoran, Phys. Rev. A {\bf 59}, 116 (1999).

\bibitem{miled3}  M. H. Y. Moussa and B. Baseia, to appear in J. Mod. Phys.
B.

\bibitem{kimble}  S. L. Braunstein and H. J. Kimble, Phys. Rev. Lett. {\bf 80%
}, 869 (1998).

\bibitem{miled1}  M. H. Y. Moussa, Phys. Rev. A {\bf 54}, 4661 (1996); M. S.
Zubairy, {\it ibid }{\bf 58}, 4368 (1998); S. Stenholm and P. J. Bardroff, 
{\it ibid }{\bf 58}, 4373 (1998).

\bibitem{barnett}  D. T. Pegg, L. S. Phillips, and S. M. Barnett, Phys. Rev.
Lett. {\bf 81}, 1604 (1998).

\bibitem{barnett1}  S. M. Barnett and D. T. Pegg, Phys. Rev. Lett. {\bf 76},
4148 (1996).

\bibitem{basilio}  B. Baseia, M. H. Y. Moussa, and V. S. Bagnato, Phys.
Lett. A {\bf 231}, 331 (1997).

\bibitem{jeffers}  S. Barnett, J. Jeffers, and A. Gatti, Phys. Rev. A {\bf 57%
}, 2134 (1998).

\bibitem{dakna}  M. Dakna, J. Clausen, L. Kn\"{o}ll, and D.-G. Welsch, Phys.
Rev. A {\bf 59}, 1658 (1999).
\end{references}
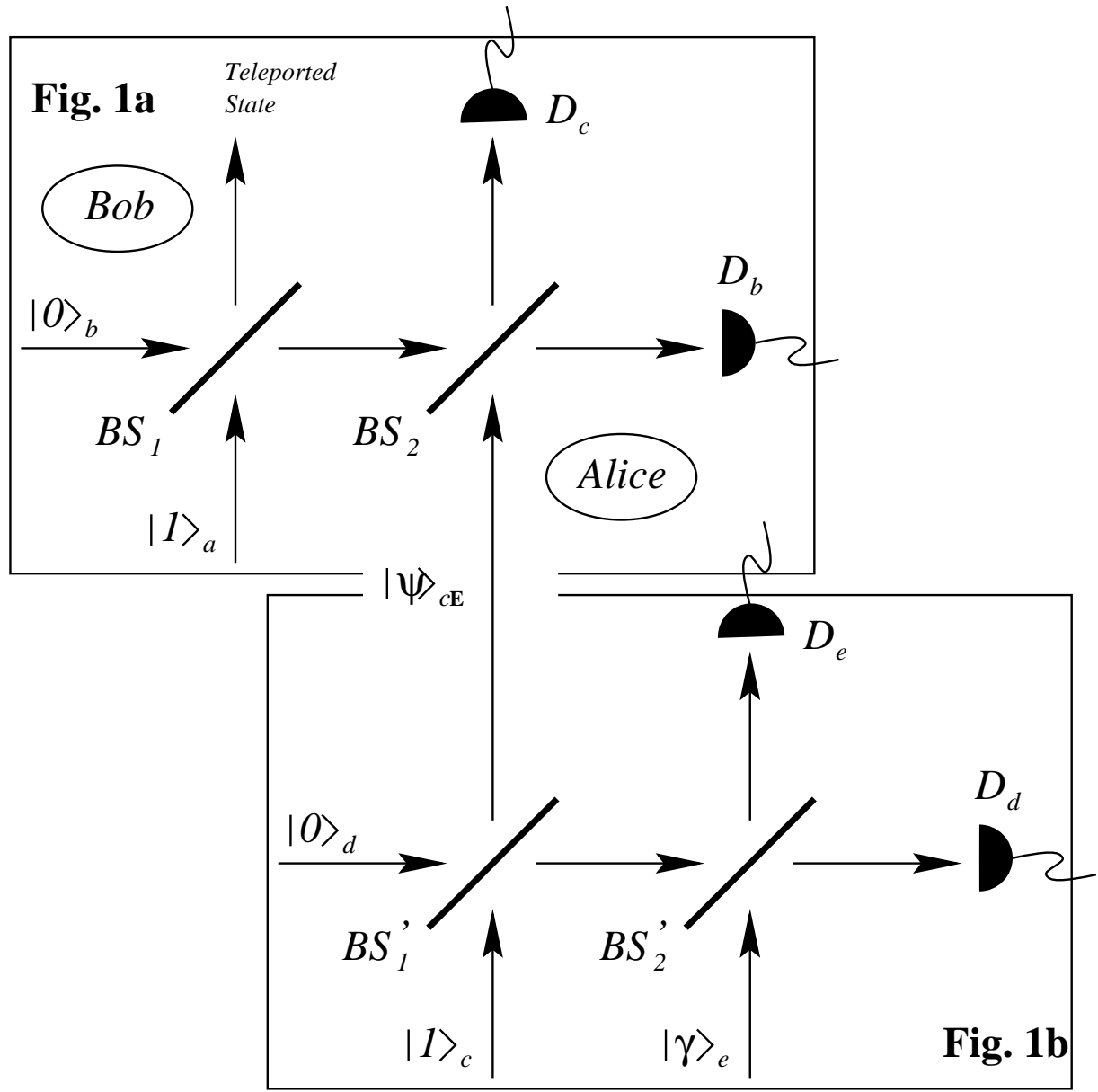
\end{document}